\documentclass[onecolumn,nobibnotes,nofootinbib]{revtex4}
\usepackage{amsmath,amssymb}
\usepackage{graphicx,subfigure,epsfig}
\newcommand{\be}{\begin{equation}}
\newcommand{\ee}{\end{equation}}
\newcommand{\bea}{\begin{eqnarray}}
\newcommand{\eea}{\end{eqnarray}}
\newcommand{\benn}{\begin{displaymath}}
\newcommand{\eenn}{\end{displaymath}}
\newcommand{\beann}{\begin{eqnarray*}}
\newcommand{\eeann}{\end{eqnarray*}}

\begin{document}

\title{Froissart bound and gluon number fluctuations}
\author{Wenchang Xiang\footnote{wxiangphys@gmail.com}}
\affiliation{Department of Physics, The University of South Dakota,
Vermillion, SD 57069, USA\\
Fakult$\ddot{a}$t f$\ddot{u}$r Physik, Universit$\ddot{a}$t
Bielefeld, D-33501 Bielefeld, Germany}
\date{\today}

\begin{abstract}
We study the effect of gluon number fluctuations (Pomeron loops) on the impact
parameter behavior of the scattering amplitude in the fixed coupling case.
We demonstrate that the dipole-hadron cross-section computed from gluon number
fluctuations saturates the Froissart bound and the growth of the radius of the
black disk with rapidity is enhanced by an additional term as compared to the
single event case. We find that the physical amplitude has a Gaussian impact
parameter dependence once the gluon number fluctuations are included. This
indicates that the fluctuations may be the microscopic origin for the Gaussian
impact parameter dependence of the scattering amplitude.

\end{abstract}

\maketitle

\section{Introduction}
\label{sec:intro} The consistent description of the impact parameter
behavior of the scattering amplitude is a long standing problem. In
this work we discuss the influence of gluon number
fluctuation on this behavior.

In fact, some activities toward understanding how fluctuations
change the impact parameter dependence of the scattering amplitude
have already started (see for example
Refs.~\cite{Iancu:2007npa,Hatta:2007npa,Hatta:2008jhep}).
Nevertheless, it is necessary to admit that we are still far away
from a complete and consistent theory related to this subject. The
most crucial difficulty related to the impact parameter dependence
is the {\it non-perturbative} (soft) contribution, which should be
taken into account at large values of the impact parameter (for
reviews see ~\cite{HM,Mueller:9911289,Levin:9710546} and references
therein). In our approach we use a different technique compared to
Refs.~\cite{Iancu:2007npa,Hatta:2007npa,Hatta:2008jhep}. Namely, we
calculate the rapidity dependence of the radius of the black disk in
the fluctuation-dominated (diffusive scaling) region at high energy.

The gluon number fluctuations become important at very high energy.
Therefore, when considering the way how the Froissart bound may
emerge based on the knowledge gathered in the small-x physics, the
effects of the most recent elements in the evolution and the effects of
Pomeron loop, have to be taken into account.

In this work we focus on the consequences of fluctuations on the
impact parameter dependence of the scattering amplitude. In
Section~\ref{subsec:BD} and~\ref{sec:bPar}, we will briefly review
the Froissart bound and the non-perturbative input of the scattering
amplitude. The Froissart bound including gluon saturation effects
will be studied in Section~\ref{cht4:sec3}. It turns out that the
total cross section saturates the Froissart bound in the case of
gluon saturation. In Section~\ref{sec:sol}, we will compute the
impact parameter dependence of the physical amplitude including
gluon number fluctuations. We find that the physical amplitude has a
Gaussian dependence on the impact parameter, which is in agreement
with experimental measurements. We also calculate the radius of the
black disk including gluon number fluctuations and find a unique
rapidity dependence coming from fluctuations. Further, in
Section~\ref{sec:phem} we calculate the slope parameter $B$.
The summary is given in Section~\ref{sec:ccl}.

\section{Unitarity and Froissart bound} \label{subsec:BD}

In high energy scattering processes, the total cross section for the
reaction $12\longrightarrow n$ particles is
\be \sigma_{12\longrightarrow
n}=\frac{1}{4\vert\bf{p_1}\vert\sqrt{s}}\sum(2\pi)^4\delta^4(P^f-P^i)
\vert\langle f_n\vert T \vert i\rangle\vert^2. \ee
Here $\vert \bf{p_1}\vert$ is the magnitude of the initial
center-of-mass frame three momentum. It is well known that the
probability conservation in the scattering processes requires the
scattering $S$ matrix to be a unitary matrix  $SS^{\dagger}=1$.
Unitarity of $S$ matrix provides a simple way to derive total cross
sections from the forward($\theta_s=0$) elastic scattering
amplitude, which is known as the optical theorem. For any
orthonormal states $\vert i\rangle$ and $\langle j\vert$, one has
\be \delta_{ji}=\langle j\vert SS^{\dagger}\vert
i\rangle=\sum_{f}\langle j\vert S\vert f\rangle\langle f\vert
S^{\dagger}\vert i\rangle \ee
where we have used the completeness relation
\be \sum_{f}\vert f\rangle\langle f\vert=1. \ee
We define the $T$ matrix as $S=1-iT$, then the unitarity condition
requires that
\be \langle j\vert T\vert i\rangle -\langle j\vert T^{\dagger}\vert
i\rangle= (2\pi)^4i\sum_{f}\delta^4(P^f-P^i)\langle j\vert
T^{\dagger}\vert f\rangle\langle f\vert T\vert i\rangle. \ee
In the case of $j=i$, where the final state is the same as the
initial state, we obtain the optical theorem
\be
\sigma_{12}^{tot}=\frac{1}{2\vert\bf{p_1}\vert\sqrt{s}}\texttt{Im}\langle
i\vert T\vert i\rangle\label{optical} \ee
with
\be 2\texttt{Im}\langle i\vert T\vert
i\rangle=\texttt{Im}A(s,t=0)=\sum_{f}(2\pi)^4i\delta^4(P^f-P^i)\vert\langle
f\vert T\vert i\rangle\vert^2, \ee
where $A(s,t)$ is the elastic scattering amplitude, $s$ and $t$ are
the center-of-mass energy squared and the momentum transfer squared,
respectively.

In high energy physics, the two particles scattering amplitude
$A(s,\cos\theta_s)$ can be expanded in the partial-wave series,
\be A(s,\cos\theta_s)=16\pi
\sum_{l=0}^{\infty}(2l+1)A_l(s)P_l(\cos\theta_s),\label{scatteringAS}
\ee
where $P_l(\cos\theta_s)$ is the Legendre polynomial of the first
kind, and $\theta_s$ is the $s$ channel scattering angle in the
center-of-mass frame. The partial-wave amplitude $A_l(s)$ can be
written in terms of a real phase shift $\delta_s$ and an inelastic
threshold $\eta_l$
\be
A_l(s)=\frac{\eta_l(s)e^{2i\delta_l(s)}-1}{2i\rho(s)},\label{Altt}
\ee
where $\rho(s)=2\vert \bf{p_1}\vert/\sqrt{s}$ with our choice of
normalization, and unitarity requires that $0<\eta_l<1$. $A_l(s)$
will be exponentially small for
\be l\geq\alpha M^{-1}\sqrt{s}\ln(s) \ee
and the scattering amplitude \eqref{scatteringAS} may be truncated
at this value~\cite{barone}. With the unitarity constrain
$0<\eta_l<1$ and \eqref{Altt}, we can get
\be \vert A_l(s)\vert=\left\vert
\frac{\eta_l(s)e^{2i\delta_l(s)}-1}{2i\rho(s)}\right\vert
\leq\frac{1}{\rho(s)}, \ee
where $\rho(s)\longrightarrow1$ as $s\longrightarrow\infty$. We know
that the Legendre polynomial of the first kind $\vert
P_l(\cos\theta_s=1)\vert\leq 1$. So, for large $s$
\be \vert A(s,\cos\theta_s=1)\vert
\leq\sum_{l=0}^{l_m}(2l+1)\label{fAS} \ee
with~\cite{Pomeronphysics}
\be l_m=\alpha M^{-1}\sqrt{s}\ln(s). \ee

Performing the summation over $l$ in \eqref{fAS} gives
\be \vert A(s,\cos\theta_s=1)\vert \leq \texttt{constat}\times s
\ln^2(s). \ee
Substituting it into \eqref{optical}, the Froissart bound
is~\cite{Froissart:1961prd}
\be \sigma^{tot}\leq\texttt{constant}\times\ln^2(s) \ee
where we have used
\be \vert{\bf p_1}\vert^2s=(P_1\cdot
P_2)^2-m_1^2m_2^2=\frac{1}{4}[s-(m_1+m_2)^2][s-(m_1-m_2)^2]. \ee
Here, $P_1$ and $P_2$ are the magnitude of the initial
center-of-mass frame four momentum in the two-body scattering
process $1+2\longrightarrow3+4$, the $m_1$ and $m_2$ are the mass of
particle $1$ and particle $2$. The Froissart bound is one of the
outstanding results of the analytic $S$ matrix theory. This bound
has been derived in 1961 by Froissart\cite{Froissart:1961prd} assuming
that the two particles scattering amplitude has uniformly bounded
partial wave amplitudes and satisfies the Mandelstam representation
with a finite number of subtractions. The Froissart bound expresses
that the hadronic total cross section can not rise faster than
$\texttt{constant}\times\ln^2s$.


\section{The non-perturbative input}  \label{sec:bPar}

In many practical applications, it is too complicated to perform
calculations of the scattering amplitude keeping precisely the
information about the impact parameter dependence, since it is
related to the {\it non-perturbative} physics. In order to simplify
the situation, one considers the scattering at fixed impact
parameter and then introduces the knowledge about the impact
parameter dependence through some profile function, which we will
denote by $S(b)$. Usually, the following two ans\"{a}tze are used as
a {\it non-perturbative} input:

\begin{enumerate}
    \item  The scattering amplitude expressed as
    the product of the scattering amplitude at fixed impact parameter times the profile
    function $S(b)$

\begin{equation} \label{PAR1}
T(Y,r,b) = T(Y,r) \cdot S(b).
\end{equation}
Such factorization form is usually used in the region of large
values of the impact parameter $b$.

\item The second one is mostly inspired by the numerical study of BK equation with
a modified BK kernel in which the kernel of the BK integral equation
is regulated to cut off infrared
singularities~\cite{Takashi:2004npa}. The impact parameter
dependence is introduced through the saturation scale, $Q_s(Y,
b)=Q_s(Y) \cdot S(b)$, and consequently for the scattering amplitude
we have:

\begin{equation} \label{PAR2}
T(Y,\,Q,\, b) \; = \; T(Q, \, Q_s(Y, b)) \; = \; T(Q, \,Q_s(Y, b=0)
\cdot S(b)) \,.
\end{equation}

\end{enumerate}

In both cases, the impact parameter profile function typically has
the exponential behavior $S(b)=e^{-2m_\pi b}$ at large distances
$b\gg R_0$, where $R_0$ is the typical radial size of the hadron
under consideration and $R_0$ increases as $A^{1/3}$ for a nucleus
with atomic number $A$. We use such an exponential fall-off at large
impact parameter as a {\it non-perturbative} initial condition at
low energy.


\section{Single event amplitude}\label{cht4:sec3}
In the geometric scaling region and in the fixed coupling case, the
dipole-hadron scattering amplitude reads \be\label{geoscaling} T(Y,r,b)\simeq
\left(r^2Q_s^2(Y)\right)^{\gamma_s}\cdot S(b), \ee where the
saturation momentum is \be Q_s^2(Y)=Q_0^2e^{\omega\bar\alpha_sY} \ee
with the arbitrary reference scale $Q_0$
($Q_0\sim\mathcal{O}(\Lambda_{QCD})$) and with the $S(b)$ giving the
impact parameter dependence. Note that here factorization is
assumed, which is the case as in
Refs.\cite{Ferreiro:2002npa,Takashi:2004npa}. Eq.~\eqref{geoscaling}
shows geometric scaling with the anomalous dimension \be
\gamma=1-\gamma_s\simeq0.37. \ee

Now, with the {\it non-perturbative} input \be S(b)\simeq
e^{-2m_{\pi}b} \ee at large $b$, one obtains from the condition \be
T(Y,r,R)=\kappa\simeq\mathcal{O}(1) \ee the ``black disc radius''
\be\label{rbdisc} R\simeq
\frac{\gamma_s}{2m_{\pi}}\left(\omega\bar\alpha_sY-\ln\left(\frac{Q^2}{Q_0^2}\right)\right).
\ee Eq.~\eqref{rbdisc} gives the standard result given in the
literature~\cite{Ferreiro:2002npa}. We have gone through such a
detailed “derivation” of~\eqref{rbdisc} since one of the main
purposes of the present work is to show how Eq.~\eqref{rbdisc} is
modified once gluon number fluctuation effects are included.

The resulting dipole-hadron cross section saturates the Froissart bound \bea
\sigma^{tot}&=&2\int d^2b T(Y,r,b)\nonumber\\
&&=2\pi R^2\nonumber\\
&&\simeq\frac{2\pi\gamma_s^2}{4m_{\pi}^2}\left(\omega\bar\alpha_sY-\ln\left(\frac{Q^2}{Q_0^2}\right)\right)^2\nonumber\\
&&\sim\frac{\pi\gamma_s^2}{2}\left(\frac{\omega\bar\alpha_s}{m_{\pi}}\right)^2\ln^2s
\eea with $Y=\ln(s/Q^2)$.

We would like to note that there was some controversy between \cite{Ferreiro:2002npa} and \cite{kovner:2002prd,kovner:2003plb}.
The authors of Refs.\cite{kovner:2002prd,kovner:2003plb} claim that the exponential fall-off with $b$ of the initial
distribution should replace by a power law fall-off due to perturbative nature of BK equation, which would be then too slow
to satisfy the Froissart bound. While Ref.\cite{Ferreiro:2002npa} points out that the BK equation will preserve the exponential
tails at very high energy due to a quasi-locality of the BK equation. The numerical study of the BK equation with a modified BK
kernel shows that the exponential tail is preserved when the kernel of the BK equation is properly regulated in the infrared\cite{Takashi:2004npa}.

It is easy to check that both ans\"{a}tze in Eq.~\eqref{PAR1} and
Eq.~\eqref{PAR2} for the single event amplitude lead to the similar result
in Eq.~\eqref{rbdisc}.

\section{Including gluon number fluctuations}\label{sec:sol}

After including fluctuations one has to distinguish between the even-by-event amplitude and the averaged (physical) amplitude. They can be explained by
considering the evolution of a hadron from $y = 0$ up to $y = Y$
which is probed by a dipole of size $r$, giving the amplitude $T(r, Y)$.
The evolution of the hadron is stochastic and leads to random gluon number
realizations inside the hadron at $Y$, corresponding to different events in
an experiment. The physical amplitude, $\bar{T}(r, Y)$, is then given
by averaging over all possible gluon number realizations/events,
$\bar{T}(r, Y) = \langle T(r, Y)\rangle$ , where $T(r, Y)$ is the amplitude
for the dipole $r$ scattering off a particular realization of the evolved hadron
at $Y$.

Based on the high energy QCD/statistical physics correspondence, we
can write\cite{munier:2005plb}
\begin{equation} \label{NEV}
\bar{T}(\rho,\rho_s(Y, b)) = \langle T(\rho,\rho_s(Y, b)) \rangle = \int d\rho_s(Y, b)
T(\rho-\rho_s(Y, b))P(\rho_s(Y, b) - \langle\rho_s(Y, b)\rangle)
\end{equation}
where we have used $\rho=\ln (Q^2/Q_0^2)$ and $\rho_s(Y, b)=\ln
(Q_s^2(Y, b)/Q_0^2)$ and we have assumed the dependence of the scattering amplitude on the impact parameter through
saturation scale\cite{munier:2008prd}. The probability distribution of $\rho_s(Y, b)$
is argued to have Gaussian form\cite{bwxiao06},
\begin{equation}
P(\rho_s(Y, b))\simeq \frac{1}{\sqrt{\pi
DY}}\exp\left[-\frac{(\rho_s(Y, b)-\langle\rho_s(Y,
b)\rangle)^2}{DY}\right]
\end{equation}
and the single scattering amplitude $T(\rho-\rho_s(Y, b))$ is
\begin{equation}\label{cht4Tevent}
    T(\rho,\rho_s(Y, b))=
    \begin{cases}
        \displaystyle{1} &
        \text{ for\,  $\rho \leq \rho_s(Y, b)$}
        \\
        \displaystyle{\exp \left[ -\gamma_s (\rho - \rho_s(Y, b)) \right]} &
        \text{ for\,  $\rho \geq \rho_s(Y, b)$}.
    \end{cases}
\end{equation}
It is easy to show that in the diffusive scaling region, $\sigma \ll
\rho-\langle \rho_s(Y,b)\rangle \ll \gamma_{s}\sigma^2 $,
\be\label{nonfactor} \langle T(Y,r,b)\rangle\simeq e^{-(\rho-\langle
\rho_s(Y,b)\rangle)^2/DY}\cdot\frac{\sqrt{DY}} {\rho-\langle
\rho_s(Y,b)\rangle} \ee where we have used $\sigma^2=DY$ with $D$
being the diffusion coefficient. Now, with the {\it non-perturbative}
input \be
\langle\rho_s(Y,b)\rangle\simeq\rho_s(Y,b)\simeq\omega\bar\alpha_sY-2m_{\pi}b,
\ee one can easily see that the exponential decrease with $b$ in the
single event case is turned into a Gaussian $b$ dependence
\bea\label{averamp1} \langle T(Y,r,b)\rangle&\simeq&\frac{\sqrt{DY}}
{\rho-\langle \rho_s(Y,b)\rangle}\cdot e^{-(\rho-\omega\bar\alpha_sY+2m_{\pi}b)^2/DY}\nonumber\\
&\propto&\frac{1}{2\sqrt{\pi}}e^{-\frac{4m_{\pi}^2b^2}{DY}}. \eea
This consequence of fluctuations seems to be supported by the
experimental
observations~\cite{Chekanov:epjc2002,chekanov:2004npb,aktas:0510016}
since, say for pp collision, \be \frac{d\sigma_{el}}{dt}\sim
e^{-B|t|} \ee which after a Fourier transform gives \be S(b)\sim
e^{-\frac{b^2}{2B}}. \ee

Second consequence of fluctuations is that the factorization is
broken, see Eq.~\eqref{nonfactor}, \be \langle T(Y,r,b)\rangle\neq
f(Y,r)\cdot S(b) \ee as compared to the single event amplitude in
Eq.~\eqref{geoscaling} where \be \langle T(Y,r,b)\rangle =
f(Y,r)\cdot S(b). \ee

Third consequence of fluctuations is that also the averaged
amplitude in Eq.~\eqref{averamp1} satisfies the Froissart bound.
Namely, from the condition $\langle
T\rangle=\kappa\simeq\mathcal{O}(1)$ (but $\kappa\leq 1$), \be
\kappa\simeq\frac{\sqrt{DY}} {\rho-\langle \rho_s(Y,R)\rangle}\cdot
e^{-(\rho-\omega\bar\alpha_sY+2m_{\pi}R)^2/DY} \ee which after
taking the logarithm on both sides reads ($\kappa'$ close to one)
\be
-\kappa'=-\frac{(\rho-\langle\rho_s(Y,R)\rangle)^2}{DY}+\ln\left(\frac{\sqrt{DY}}{\rho-\langle\rho_s(Y,R)\rangle}\right)
\ee and is fullfilled if \be
\rho-\langle\rho_s(Y,R)\rangle\simeq c\sqrt{DY} \ee with the coefficient $c$ of order $\sqrt{\kappa'}$.

Now, with
$\langle\rho_s(Y,R)\rangle\simeq \omega\bar\alpha_sY-2m_{\pi}R$, one
obtains \bea
\rho-\omega\bar\alpha_sY+2m_{\pi}R &=& c\sqrt{DY}\nonumber\\
\Longrightarrow
R&=&\frac{1}{2m_{\pi}}\left(\omega\bar\alpha_sY+c\sqrt{DY}-\ln\left(\frac{Q^2}{Q_0^2}\right)\right).
\eea As compared to Eq.~\eqref{rbdisc}, this equation taking
fluctuations into account contains a new term $\sqrt{DY}$.

So, including fluctuations and the impact parameter dependence in
the way presented here seem to lead to reasonable results. However,
the whole discussion is valid only in the fixed coupling case. We have
already calculated the dipole-hadron scattering amplitude by considering
both gluon number fluctuations and runing coupling effect\cite{xiangprd09};
however we do not know how to take into account the impact parameter
dependence of the scattering on top of gluon number fluctuations and
runing coupling effect. This will be considered in our next work.

\section{Phenomenological applications and estimation of the slope parameter $B$}\label{sec:phem}
It is well known from numerous hadronic scattering experiments that
\begin{equation} \label{DF1}
\frac{d \sigma_{exp}}{d t} \,\, \propto \,\, e^{- B |t|}.
\end{equation}
where $t$ is the squared four momentum transfer between the
projectile and target. The $t-$slope $B$ tends to a universal value
determined by the proton shape alone~\cite{Kowalski:2006prd}. From
the experimental measurement of the $t-$distribution of the vector
mesons, the effective slope $B$ is found to be
$B=4GeV^{-2}$~\cite{Chekanov:epjc2002,chekanov:2004npb,aktas:0510016}.

In order to study the $t-$slope $B$, we take the Fourier transform
of Eq.(\ref{DF1}):
\begin{equation} \label{DF2}
S_{exp}(b) \,\, \propto \,\, \frac{1}{2 \, \pi \, B} \,\, e^{-
\frac{b^2}{2B}} \, .
\end{equation}
Now from comparison of factors in the exponent of
Eq.(\ref{averamp1}) with Eq.(\ref{DF2}), we can immediately see that
\bea \label{DF3}
B&=&\frac{\sigma^2}{8m_{\pi}^2}\nonumber\\
&=&\frac{DY}{8m_{\pi}^2}\nonumber\\
&=&\frac{D}{8m_{\pi}^2}
\ln\left(\frac{s}{Q^2}\right) \eea where $D$ is the diffusion
coefficient. The value of $D = 0.325$ is determined by fitting the
HERA data with the color glass condensate model plus the gluon number
fluctuations~\cite{XW,Xiang:2009npa}. Note that the $B$ increases logarithmically
with the center of mass energy $s$, which is in agreement with the
Regge theory. This is the phenomenon known as the $shrinkage$ of the
diffraction peek in Regge theory, which can be interpreted as an
increase of the interaction radius $R_{int}\sim\sqrt{\ln s}$. With
the reasonable values $m_{\pi}=0.14GeV$,
$\sqrt{s}=90GeV$ and $Q^2=10GeV^2$, one obtains
values~\cite{Chekanov:epjc2002,chekanov:2004npb,aktas:0510016} \bea
B&=&\frac{D}{8m_{\pi}^2}
\ln\left(\frac{s}{Q^2}\right)\nonumber\\
&\simeq&13.8GeV^{-2}. \eea

\section{Summary}\label{sec:ccl}

The main results of this work can be summarized as follows: We have
argued that the impact parameter behavior of the scattering
amplitude in the presence of fluctuations has Gaussian-like
behavior. Such behavior is in agreement with various
phenomenological models. This indicates that fluctuations may be the
microscopic origin for the Gaussian behavior.

Further, we have shown that the factorization of the impact parameter
of the scattering amplitude is lost once
the gluon number fluctuations are included.

We calculated the rapidity dependence of the radius of the black disk in the
fluctuation-dominated (diffusive scaling) region at high energy. We
found that, due to fluctuations, the growth of the radius of the black
disk is enhanced by an additional (proportional to square root of
rapidity) term.


\begin{acknowledgments}
I would like to thank Dr. Arif.~Shoshi for suggesting this work and
numerous stimulating discussions. Without his patient guidance, this
work would not be possible. I acknowledge Edmond ~Iancu for
discussions on these and related topics.
\end{acknowledgments}

%


\end{document}